# The Potential Energy Hotspot:
# Effects from Impact Velocity, Defect Geometry, and Crystallographic Orientation

Brenden W. Hamilton[1], Matthew P. Kroonblawd[2], Alejandro Strachan[1]*

[1]School of Materials Engineering and Birck Nanotechnology Center, Purdue University, West Lafayette, Indiana, 47907 USA
[2]Physical and Life Sciences Directorate, Lawrence Livermore National Laboratory, Livermore, California 94550, United States
strachan@purdue.edu

## Abstract

In energetic materials, the localization of energy into "hotspots" when a shock wave interacts with the material's microstructure is known to dictate the initiation of chemical reactions and detonation. Recent results have shown that, following the shock-induced collapse of pores with circular cross-sections, more energy is localized as internal potential energy (PE) than can be inferred from the kinetic energy (KE) distribution. This leads to a complex thermo-mechanical state that is typically overlooked. The mechanisms associated with pore collapse and hotspot formation and the resulting energy localization are known to be highly dependent on material properties, especially its ability to deform plastically and alleviate strain energy, as well as the size and shape of the pore. Therefore, we use molecular dynamics simulations to characterize shock-induced pore collapse and the subsequent formation of hotspots in TATB, a highly anisotropic molecular crystal, for various defect shapes, shock strengths and crystallographic orientations. We find that the the localization of energy as PE is consistently higher and its extent larger than as localized as KE. A detailed analysis of the MD trajectories reveal the underlying molecular process that govern the effect of orientation and pore shape on the resulting hotspots. We find that the regions of highest PE for a given KE relate to not the impact front of the collapse, but the areas of maximum plastic deformation, while KE is maximized at the point of impact. A comparison with previous results in HMX reveal less energy localization in TATB which could be a contributing factor to its insensitivity.

# 1. Introduction

Shockwave-induced chemistry can result in a myriad of processes such as detonation[1–3], the formation of pre-biotic compounds that may have contributed to the formation of life on Earth,[4–8] and the synthesis of new materials and phases[9–11]. Often, shock-induced chemistry is triggered or enhanced by energy localization into hotspots that form as the shockwave interacts with the material's microstructure[12]. In the case of energetic materials, hotspots of sufficient size and temperature can become critical and transition into deflagration waves and eventually lead to detonation. Several mechanisms can result in the formation of hotspots, but the collapse of porosity is known to dominate the initiation of energetic materials. This was first shown through shock desensitization experiments were high explosives (HEs) were rendered insensitive after an initial weak shock caused the collapse of porosity without igniting significant amounts of material.[13] The inclusion of inhomogeneities via silica micro-beads and cavities via micro-balloons in gelled nitromethane also demonstrated the superiority of the latter in triggering detonations, decreasing the run to detonation.[14]

Significant efforts have been devoted to understanding the formation, nature, and criticality of hotspots. Physics-based scaling laws for planar void collapse supported by atomistic simulations predict a theoretical maximum temperature achieved during pore collapse and highlight the importance of material expansion into the void, maximizing pressure-volume work during recompression.[15] Recent molecular dynamics (MD) simulations have shown that diamond shaped voids, elongated along the shock direction, result in larger and hotter hotspots than equiaxed pores.[16] This is due to focusing of shockwave energy at the tip of the diamond, leading to molecular jetting and the formation of a low density expanding plume. Volumetric work done to recompress the plume achieves temperature values close to the maximum predicted in Ref. 15. Three-dimensional calculations of the collapse of spherical and octahedron-shaped pores in HMX, showed only a nominal difference in temperature[17], however pores were limited to 8nm in the shock direction which would limit molecular jetting[16]. Continuum modeling techniques have been used to explore pore aspect ratio[18] and the resultant shear banding[19] in HMX pore collapse simulations, and compared HMX and TATB[20]. In recent years, the computational efficiency of all-atom simulations has enabled direct scale bridging with grain-scale models, opening new routes to parameterize and validate the accuracy of those models for predicting shock induced pore collapse.[21–23]

Atomic-level understanding of shock induced chemistry was greatly increased by the development of reactive force fields, such as ReaxFF, which allowed for explicit simulation of shock ignition and thermal decomposition in solid HEs such as RDX.[24,25] ReaxFF simulations using a compressive shear protocol have been utilized to explore the interplays of mechanics on chemistry.[26,27] Reactive MD techniques such as density functional tight binding have been used to explore the chemical reactivity of TATB under thermal and shock loading, as well as mechanical shear induced metallization.[28–30] The extended timescales of these techniques have allowed for the prediction of reactive properties like detonation velocity and pressure[31,32], IR spectra evolution[33–35], and detailed chemical reaction pathways[36–39]. Reactive force fields have also enabled explicit simulation of nanoscale hotspots[40] and the upscaling of chemical reaction models for mesoscale and coarse grained simulations.[41]

Quite surprisingly, reactive MD simulations have shown that nanoscale hotspots formed following the dynamical collapse of porosity are markedly more reactive than otherwise identical hotspots at the same temperature and pressure in compressed perfect crystal.[42,43] A possible explanation for this observation is that disorder and amorphization in molecular crystals can lead to accelerated reaction compared to bulk crystalline materials[44,40]. Recent MD simulations of shock-induced pore collapse in TATB showed that significantly more energy is localized as intra-molecular potential energy (PE) than into internal temperature or kinetic energy (KE).[45] This excess PE is the result of large intra-molecular deformations that do not significantly relax on timescales comparable to the onset of exothermic chemistry. Molecular deformations such as these can lead to mechanochemical acceleration of reactions and alter reaction pathways.[46] Recent work in RDX combining planar pore collapse with an edditional shear component directly linked hotspot criticality to the level of shear loading.[47] Excess localized PE provides a plausible explaination to the puzzling difference in reactivity between dynamically and thermally generated hotspots[42] and for chemical activation through forming nanoscale shear bands.[44] The generality of this observation is not well understood, nor is it understood how PE localization depends on the intrinsic properties of the crystal, the shape and size of the pore, or shock strength. Answering these questions is important to broadly characterize PE localization in hotspots and its effect on the kinetics of decomposition reactions.

To address this gap in knowledge, we characterize how different pore collapse mechanisms operating at various shock strengths (e.g. viscoplastic, hydrodynamic, molecular jetting) impacts the relative intensity and shape of the hotspot as well as the partitioning of the localized energy into kinetic (temperature) and potential (molecular strain) terms. We focus here on hot spots in the insensitive HE TATB, as its layered structure[48] leads to what is perhaps the greatest mechanical and thermal anisotropy for any explosive. This enables us to explore bounding cases for the role of shock orientation on the formation of hotspots. Recent work from Lafourcade et al. showed a strong orientation dependence for deformation mechanisms in TATB under controlled strain conditions[49] that leads to analogous deformations under shock conditions[50]. For instance, compressive stresses along [100] result in a chevron-like buckling of the basal planes, whereas compressive shear stresses along (011)-type planes results in a 'non-basal gliding' of the planes. Under weak stresses, the TATB crystal layers will glide in-plane[49,51,52] while detonation-level shocks lead to the formation of nanoscale shear bands[44]. These observations indicate that pore collapse could exhibit a high degree of effects from anisotropy as well as shock strength.

The role of TATB anisotropy in shock loading response of the perfect single crystal has been well characterized for a shock strength near 10 GPa. All-atom simulations were used to study the perfect crystal shock response in a variety of crystallographic orientations.[50] This showed significant effects on the wave structure (single vs 2-wave response), elastic wave speeds, and deformation mechanisms, which ranged from a variety of crystal level defect formations to plasticity and intense shear localization. Coupled MD and continuum simulations explored the mechanics of pore collapse for various orientations and shock speeds for cylindrical pores.[21] Strong disparity between the all-atom and the isotropic, elastic-plastic continuum models at low speeds highlighted the significance of anisotropic effects in the formation of hotspots.

Our previous work characterized the role of pore shape, size, and shock strength in hotspot formation in HMX.[16] We use identical geometries here to enable a direct comparison between TATB (considered an insensitive explosive) and HMX (a high performance material). The extreme temperatures found in HMX following the collapse of diamond-like cracks elongated along the shock propagation direction (>5000 K) corresponds well to the theoretical maximum temperature[15] and recent experimental reactive hotspot measurements from Bassett et. al.[53–55] These high temperatures are possibly related to the jetting and gasification of material into the void, which is later recompressed by the shockwave. Holian et. al. showed for simple 1D shocks in a model system that jetting occurs when the energy embedded by the shock is greater than the crystals cohesive energy: $\frac{1}{2}mU_p^2 > E_{coh}$[15]. However, events such as plasticity, dislocation formation, and shock focusing at curved surfaces can alter the local energy created from shock compression.

This paper assesses the localization of energy, both kinetic and potential, in TATB following the shock-induced collapse of porosity. The use of two void shapes allows us to evaluate the role of molecular jetting, whereas the two crystallographic orientations used elucidates the role of molecular/crystal-level processes involved in hotspot formation. We find that TATB follows the general trends observed in HMX[16] in terms of shock strength and pore shape, but with an important quantitative difference. Unlike in HMX, the temperatures achieved in TATB are only a fraction of the theoretical maximum. Our atomistic simulations provide insight about the underlying molecular processes behind this observation, which may be a contributing factor to the insensitivity of TATB. We find strong correlations between crystal orientation and hotspot shape and size. These effects can be understood from the underlying crystal processes. Finally, we find that in all cases more energy in hotspots is stored as PE than as KE.

## 2. Methods

MD simulations were conducted using the LAMMPS package[56] and a validated version of a non-reactive, non-polarizable force field for TATB.[57] The force field includes tailored harmonic bond stretch and angle bend terms for flexible molecules[58], RATTLE constraints that fix the N-H bonds to their equilibrium values,[59] and an intramolecular O-H repulsion term that was implemented as a bonded interaction.[60] The covalent bond vibrations, angle bends, and improper dihedrals were modeled using harmonic functions. Proper dihedrals were modeled using a cosine series. Van der Waals interactions were modeled using the Buckingham potential (exponential repulsion and a $r^{-6}$ attractive term) combined with short-ranged $r^{-12}$ potentials that compensate for the divergence in the Buckingham potential at small separation. The non-bonded terms were evaluated in real space within an 11 Å cutoff. Electrostatic interactions were calculated between constant partial charges located on the nuclei and were evaluated using the short-ranged Wolf potential with a damping parameter of 0.2 Å$^{-1}$ and an 11 Å cutoff.[61] All intra-molecular non-bonded interactions are excluded by design, which allows for rigorous separation of inter- and intra-molecular potential energy terms.

Nearly orthorhombic simulation cells were prepared using the generalized crystal-cutting method[62] starting from the triclinic $P\bar{1}$ TATB crystal structure[48] with lattice parameters determined with the TATB FF at 300 K and 1 atm. For orientations denoted as (001), the crystal was oriented such that [100] was aligned with x, [120] was nearly parallel to y, and the normal to the basal planes $\mathbf{N}_{(001)} = \mathbf{a} \times \mathbf{b}$ was aligned with z, the shock direction. For the (100) orientation, $N_{(100)} = \mathbf{b} \times \mathbf{c}$ was aligned with z, and the x axis was aligned with [001] (lattice vector c). For cylindrical pores, a diameter of 40 nm was used with a through axis along x, centering the void in the geometric center of a cell. For diamond-like crack-shaped pores, a diamond was cut with the long axis aligned with the shock direction (z) and the short axis aligned with the y direction (the simulation is thin in the x direction). The crack length was 40 nm and its maximum width was 8 nm. Renderings of both defect shapes and the utilized crystallographic orientations are displayed in Figure 1.

Free surfaces were generated normal to the shock direction (z) by adding a 5 nm region of vacuum that removes the periodicity in that direction to prevent self-interactions. Periodic boundaries were utilized in both non-shock directions. The thermalized systems were equilibrated at 300 K using a 25 ps isothermal-isochoric (NVT ensemble) simulation with a Nose-Hoover-style thermostat and a 0.5 fs timestep.[63] To accelerate the equilibration of the system after the free surfaces were created, during the first 2.5 ps, atomic velocities were re-initialized stochastically from the Maxwell-Boltzmann distribution every 0.5 ps and were rescaled to the target temperature every 0.05 ps to attenuate breathing modes incurred by the surface tension. These configurations were used as the starting point for reverse ballistic shock simulations using adiabatic MD (NVE ensemble) with a 0.2 fs timestep. In the reverse ballistic setup,[64] the piston velocity, $U_p$, was added to the thermal velocities of the atoms leading to impact on the rigid piston that generates a shock front traveling through the sample in the opposite direction at the shock speed, $U_s$. Molecules with center of mass positions with z ≤ 1.5 nm were held fixed throughout the shock simulation to simulate the rigid and infinitely massive piston. We ran shock simulations at $U_p$ = 0.5, 1.0, 1.5, and 2.0 km/s for each pore shape and crystal orientation case yielding a total of 16 simulations.

Simulation trajectories were analyzed on a molecule-by-molecule basis. The molecular center of mass (CM) positions and velocities were computed as weighted sums over all 24 atoms in each molecule. The total molecular kinetic energy $KE_{tot}$, and the separate contributions from the molecular translational $KE_{trans}$, and roto-librational and vibrational $KE_{ro\text{-}vib}$ degrees of freedom were computed as

$$K_{tot} = \sum \frac{1}{2} m_i \mathbf{v}_i \cdot \mathbf{v}_i$$

and

$$K_{trans} = \frac{1}{2} M \mathbf{V} \cdot \mathbf{V}$$

and

$$K_{ro-vib} = K_{tot} - K_{trans}$$

where lowercase variables represent mass and velocity of individual atoms and capital letter represent CM (molecular) values. The ro-vib kinetic energies $KE_{ro\text{-}vib}$ were interpreted as the molecular temperature T and were scaled to Kelvin units through

$$K_{ro-vib} = \frac{63}{2} k_B T$$

where $k_B$ is the Boltzmann constant and the factor of 63 arises from the 3 roto-librational and 60 unconstrained vibrational degrees of freedom in the TATB molecule. All molecular properties were locally averaged within a sphere 1.5 nm in radius about each molecule CM to smooth fluctuations.

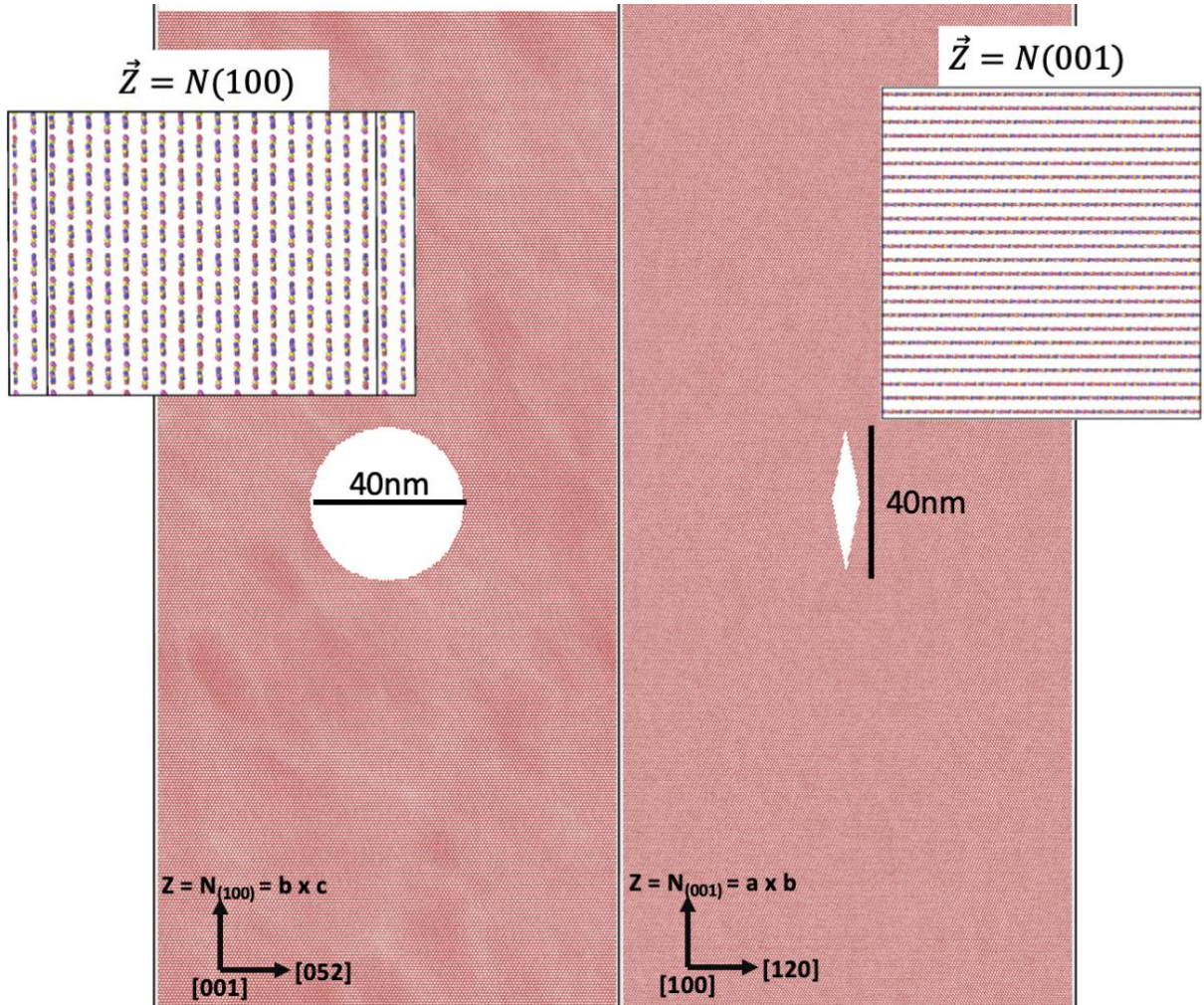

*Figure 1: Simulation set up for shock interaction with pores (left) and cracks elongated along the shock direction (left), inset figures display the two crystallographic orientations studied both for pores and cracks. Shocks propagate from bottom to top, with periodic boundary conditions in the other two directions.*

## 3. Temperature and Potential Energy Fields of Hotspots

We performed shock simulations with $U_p$ ranging from 0.5 to 2.0 km/s. Figure 2 shows molecular renderings for the various hotspots generated after pore collapse colored by both KE (left) and PE (right), for $U_p$ =1.0 to $U_p$ =2.0 km/s. The 0.5 km/s results are omitted from this post-collapse analysis as the energy difference in the KE and PE hotspots was smaller than thermal fluctuations and the crack cases produced almost no hotspot at all. Note that the color bar upper bound depends on $U_p$ and is 100, 75, and 50 kcal/mol for the 2.0, 1.5, and 1.0 km/s cases, respectively. Each pair of columns collects a shock speed, with the left panels of each group showing temperature (in units of KE) and the right half showing intra-molecular PE. Each row corresponds to an orientation and defect pair. As will be discussed in detail below, we find that for all cases with $u_p \geq 1$ km/s, the highest PE value is larger than the corresponding KE value and more energy overall is stored as PE than as KE. This confirms that the result in Ref. 45 is quite generally applicable to TATB shocks.

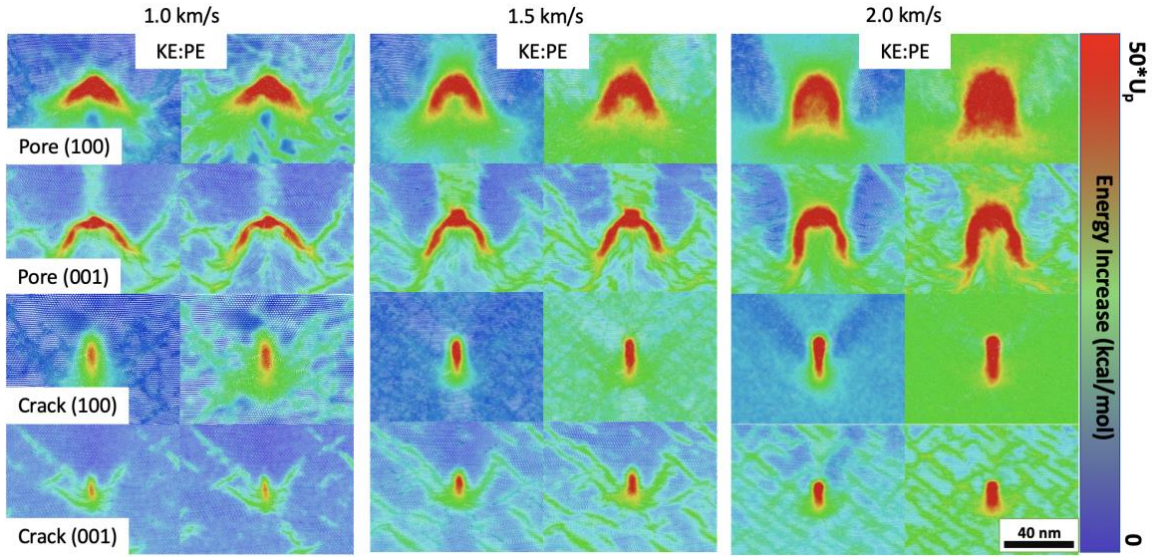

*Figure 2: Molecular renderings of all hotspots at 5ps after total collapse of porosity, colored as both kinetic energy (which is proportional to temperature) and intra-molecular potential energy. Color bar relative to impact velocity (max value 50, 75, and 100 kcal/mol, for 1, 1.5, and 2 km/s respectively).*

We note that the significant anisotropy of TATB is manifested in the markedly different shock-induced plasticity and the associated dissipation away from the hotspot areas, see Figure 2. This, together with the anisotropic elasticity, results in significantly higher bulk temperatures for (001) shocks as compared to (100). For example, a (001) shock with $U_p$ = 2.0 km/s has a shock velocity of 6.2 km/s and results in a temperature increase of ~770 K, whereas a (100) shock with the same particle velocity has a $U_s$ of 7.0 km/s and an average temperature of ~650 K, which is consistent with those found in Ref 21. In addition, (100) shocks lead to relatively homogeneous temperature fields in the bulk, whereas the (001) orientation localizes energy in shear bands.

## 3.1 Role of shock orientation

The potency of a hotspot is releated to both its size and temperature, since the critical temperature for ignition decreases with size[65]. Thus, to quantify the thermal fields of hotspots we compute the area (A) of the hotspot with temperature exceeding a value temperature (T) and plot this relationship in the T-A space. Figure 3 shows the T-A plots for cylindrical pores (3a) and diamond-like cracks (3b) for various $U_p$ and shock orientations; solid lines indicate (001) shocks while dotted lines denote (100) shocks. To single out the rise in temperature from the collapse of porosity, we reference the temperature field to the value corresponding bulk shock temperature, which may include heating from shear band formation in strong shock cases. We find that the collapse of cracks results in smaller and colder hotspots than that of pores, except for the strongest (100) shocks where the maximum temperature resulting from the collapse of cracks is over 500 K higher than in the case of pores. The molecular mechanisms behind this observation will be discussed in Section 4.

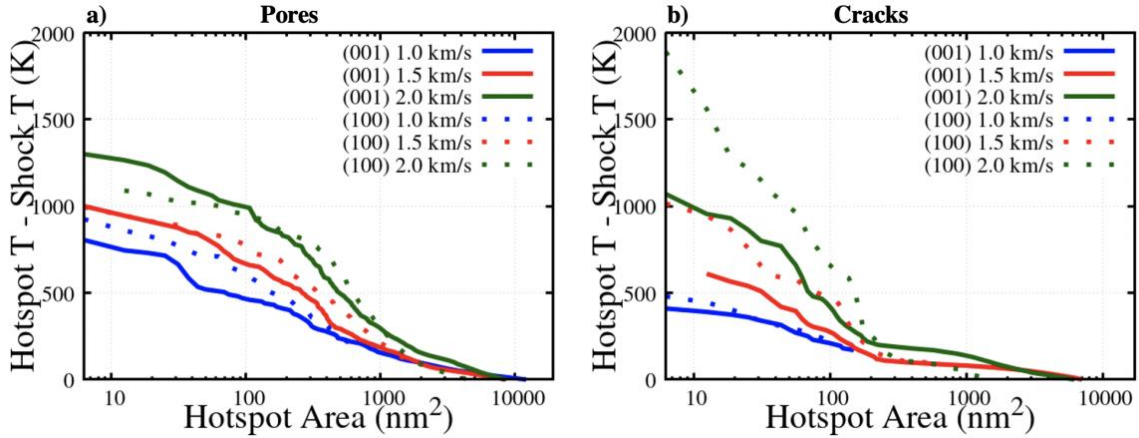

Figure 3: Temperature- Area cumulative plots for pores (a - cylinder voids) and cracks (b - diamond voids). Impact speed shown by color, orientation shown by point shape.

Starting with the two cylindrical pores at 2 km/s and assessing the role of shock direction, it is evident that, despite their similar temperature-cumulative area plots (Figure 3a), the temperature fields command highly dissimilar shapes (compare to Figure 2). The (100) shock direction pore collapse results in a rather equiaxed hotspot, indicative of the initial cylindrical void. However, the (001) shocks result in a crescent-shaped hotspot, with a discernable 'core' and 'legs'. These observations also apply to the PE component of the energy localization, as well as for lower shock speeds. The crystal scale processes (e.g., intense shearing and plasticity) that create these differing shapes will be discussed in Section 4. While the maximum temperatures and extent of the energy localization are similar for the two shock directions (see Figure 3a), differences in the initial temperature fields and post-shock densities can result in different thermal dissipation rates[66] and thus may exhibit different thresholds for reaction/deflagration. Thus, the criticality of the hotspots following the collapse of such pores can be expected be dependent on shock direction.

Finally, while the collapse of circular pores results in similar T-A distributions for both orientations (Figure 3a shows differences not exceeding 100 K for nearly all areas), the collapse of cracks results in highly anisotropic results, with the crack in the (100) shock case at 2.0 km/s reaching much higher peak temperatures.

## 3.2 Role of shape: pores vs. cracks

In the case of cracks, strong shocks ($U_p$=2.0 km/s) result in higher hotspot temperatures than the corresponding pore but command smaller areas. This difference in area is due to the smaller initial area of the defect. For 1.5 km/s shocks, cracks and pores result in similar maximum hotspot temperatures and for weaker shocks the pores result in hotter hotspots. These same trends were also observed for HMX in Ref 16. This is because in the viscoplastic regime corresponding to weaker shocks, where collapse is driven by the compressive stresses caused by the shock, the cracks along the shock direction collapse laterally and this process involves little plastic deformation. The pores, in contrast, undergo significant plastic flow during collapse from a weak shock wave due to their larger areas. For higher shock strengths, we observe fast expansion and jetting in the cracks due to shock focusing, which results in higher temperatures than for the circular pores that experience a hydrodynamic collapse. In terms of shock orientation, (100) shocks interacting with cracks elongated along the shock direction result in hotter and larger hotspots than the (001) shocks, see Figures 2 and 3(b).

Figure 4 compares the total hotspot temperature (not referenced by bulk shock temperature) vs. area distributions for both orientations and defect shapes for $U_p$ = 2.0 km/s. In this hydrodynamic regime, the peak temperature is highest for the (100) shock interacting with a crack, but only surpassing those of the pore collapsed by the (001) shock in very small areas, i.e. at the core of the hotspots. Regarding the large difference in maximum temperatures between the two crack systems, it was shown from previous work on HMX[16] that the efficiency of the cracks results from molecular spall and jetting during collapse.

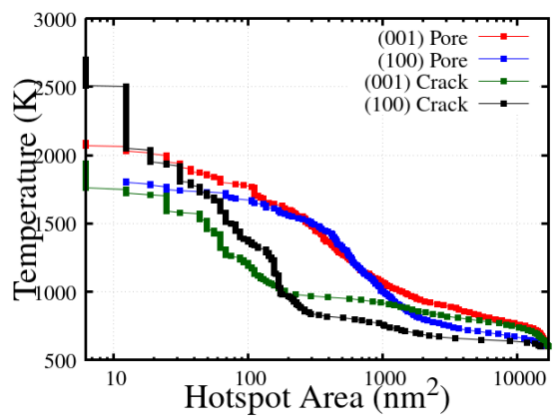

*Figure 4: T-A plot (absolute temperature) for both void shapes and crystal orientations, at 2.0 km/s impact speed.*

## 3.3 Localization of potential energy

Figure 5 shows PE vs cumulative Area (PE-A) plots for intra-molecular potential energy that are analogous to the T-A plots in Figure 3. Here we extend the analysis from Ref. 45, in which only a pore under a (001) shock was studied for $U_p$ = 2.0 km/s. We find that the trends described above for the temperature fields (in terms of role of shock strength, orientation, and pore shape) also apply to the localization of PE. The maximum PE following the collapse of a pore by a (001) shock is greater than for a (100) shock, so much so that the 1.5 km/s (001) shock surpasses the 2 km/s (100) shock at very small areas (Figure 5a). Importantly, we showed in Ref. 45 that the temperature field does not uniquely determine the PE field. That is, a one-to-one mapping between local T and PE is not possible. For example, the T-A curve following the collapse of a pore by a 1.5 km/s (100) shock is at slightly higher than that for the (001) shock at small areas, and considerably higher at larger areas, see Fig. 3. However, this trend reverses in terms of PE, with the (001) shock pore collapse leading to a slightly "hotter" hotspot in PE terms for all areas. This is indicative of a loading path dependence in how hotspot energy partitions between KE and PE. In Section 5, we will more closely inspect the PE-T distributions for all cases.

The T-A and PE-A trends are relatively closer in the case of cracks; however, the 2.0 km/s (100) crack has a significantly larger increase of PE over the KE, relative to all other shocks. In the next section, we will assess the crystal-level processes that occur during the collapse of the void and relate the orientation effects to the relevant hotspot results shown here.

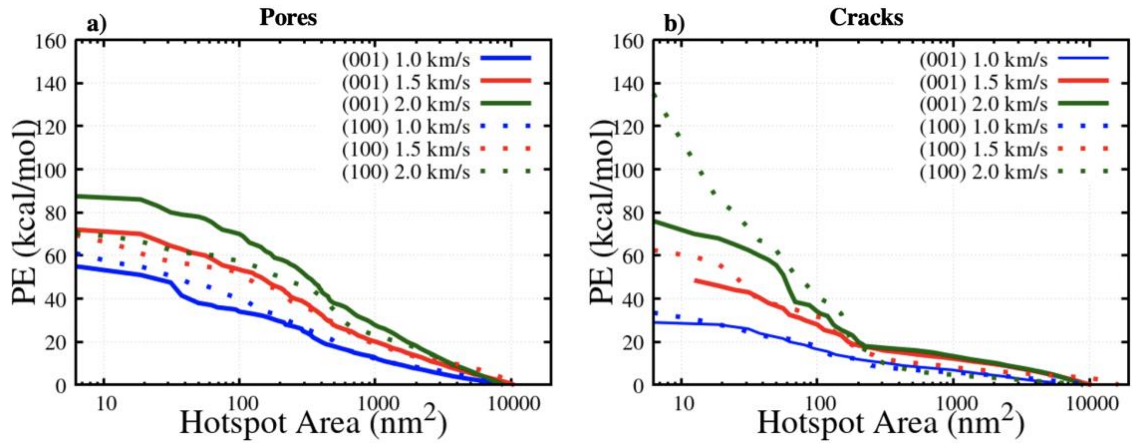

Figure 5: PE-A plot for cracks and pores. Organized the same as Figure 2, except the y axis correspond to 0-1400 and 0-2200 in Temperature units, respectively.

## 4. Crystal-Level Processes of Pore Collapse

To better understand the formation of hotspots following the collapse of porosity we now characterize the process of collapse at molecular and crystal-level scales. Starting with cylindrical pores, see Figure 6, the general shape of the collapsing pores only shows minor orientation dependence, but, as expected, strong dependence on shock strength. For $U_p$=0.5 km/s, cylindrical voids collapse laterally via a viscoplastic process driven by the compressive stresses following the passage of the shock. With increasing shock strength, the collapse transitions to a hydrodynamic regime where the upstream surface expands into the void with timescales similar to the passage of the shock. In this regime, the stresses involved are significantly higher than the material strength[49,67] and the deforming material behaves approximately like a fluid (hence the name hydrodynamic).

As could be expected, the collapsing material under (100) shocks lacks long-range order, it is amorphized by the fast plastic deformation. Interestingly, for (001) shocks, the collapsing material retains a significant degree of crystalline order, see the second row of Figure 6. In this case, the TATB basal planes are highly deformed via layer sliding and non-basal gliding as well as plane buckling but remain otherwise crystalline. This retained crystallinity is likely related to the small number of slip systems available for plastic deformation[49] in conjunction with dislocation motion instability under shock-like pressures[68] that leads to nanoscale shear banding[44,50] as a primary plastic response for this orientation.

To further characterize the structure of the collapsing material, Figure 7 displays the 3-dimensional radial distribution functions obtained from the molecular centers of mass for both collapsing regions. This clearly confirms the structural difference in the collapsing material. While both orientations result in mostly amorphous material in the hotspot after the collapse and recompression (Fig. 7(b)), the collapsing material is crystalline for the (001) and amorphous for the (100), Fig. 7(a). This difference in the local structure of the collapsing material explains the difference in the resulting shape of the hotspots displayed in Figure 2. The higher degree of crystallinity in the (001) case results in more localized deformation along the downstream surface perimeter and, consequently, a hotspot with a core at the point of collapse and legs marking the regions of intense shear. The amorphous material in the (100) pore collapse leads to a more uniform, circular hotspot, as compared to the (001) pore. Additionally, the amorphization of the material during collapse in the (100) shocks may allow for more intra-molecular relaxations. This may explain why the (001) shocks with pores lead to hotter hotspots in the PE description than the (100) shocks, relative to the KE hotspots they form. The collapsing material is at low density and will allow individual molecules more freedom to conformationally relax when in the amorphous state, while the crystalline molecules of the (001) are more "locked into place" prior to total collapse of the void.

We now turn our attention to orientation effects in the shock-induced collapse of cracks where we will again focus on the $U_p$ = 2.0 km/s case. Ref. 16 showed that the extreme temperatures following the collapse of cracks were a result of molecular jetting in HMX. Without access to the results shown in Figure 6, one would unwittingly assume that the crack collapse driven by (100) shocks are hotter than (001) due to increased jetting, and that anisotropic elasticity and plasticity[49] would result in different amounts of jetting. However, upon close inspection of Figure 6, the cracks along the (001) and (100) shock directions exhibit similar amounts of ejecta at early times, which agrees well with the simple relationship put forward by Holian et al., which states that jetting occurs when the shock energy is greater than the crystal's cohesive energy, $\frac{1}{2}mU_p^2 > E_{coh}$.[15] In fact, the key difference between the two cases, see Figure 6, is the location of the shock front (transition from green to blue coloring) relative to the ejecta. The particle velocity of the ejecta relative the shock speed is higher for the (100) case; this is true even though the shock speed of the (100) case is higher than in the (001) case ($U_s$ of 7.02 km/s vs. 6.23 km/s[21]). Thus, a (100) shock interacting with a crack ejects material out in front of the shock, the ejecta expands over the void unimpeded and recompresses in the downstream end of the defect. In the (001) case, the compression caused by the shock front closes the crack ahead of the ejecta limiting expansion and resulting in lower temperatures. The remaining challenge is to understand the relative velocities between ejecta and shock speed for the two cases.

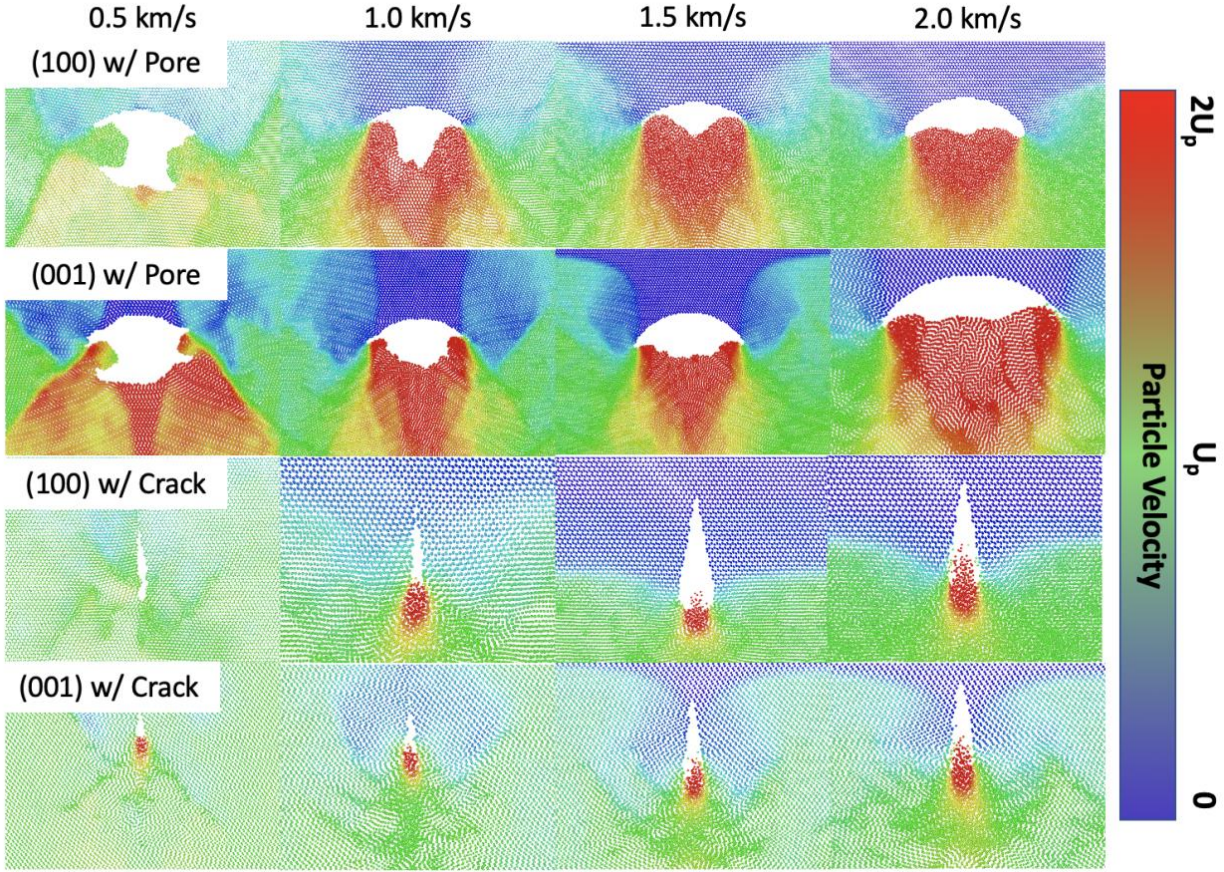

*Figure 6: Molecular COM velocities in the shock direction during pore collapse. Color bar relative to initial impact velocity.*

When a shockwave reaches a flat free surface, the material expanding into vacuum travels at $2U_p$, or double the incoming particle velocity.[69] In the case of nonplanar defects, shock focusing can lead to much higher ejecta velocities. When a (001) shock interacts with a crack, the ejected molecules have velocities roughly between 4 and 6 km/s, whereas the shock along (100) results in molecular velocities mainly between 5 and 7 km/s, allowing the ejecta to expand quicker (see Figure 12 in Section 6 for full velocity distributions). Additionally, the (001) shock case has a distinct two wave feature in which the leading wave results in significant pressure rise. This can initiate lateral compression of the crack but does not have the energy to eject material into the crack, except for the 2.0 km/s case, in which the leading wave is about ~0.3 km/s above the threshold as defined by Ref. 15. Wave profiles for both orientations are shown in Figure 8. As the shock travels along the crack, the compressive stresses result in lateral collapse and can begin to occur before jetting takes place. In our previous work on HMX[16], not all cracks that exhibit jetting resulted in high hotspot temperatures. In cases where the jetting was slow relative to the shock, the crack could laterally collapse before the ejecta impact on the downstream face, trapping the ejecta short of their goal and preventing the excess volumetric work created from gaseous expansion and recompression. This lateral collapse limits the ejecta in the (001) case since these relaxations will begin to occur prior to full expansion of the ejected material, choking off the ejecta before recompression. In the (100) case, the ejecta is accelerated in front of the shock wave, and its rapid free expansion will be unaffected by the comparatively slow lateral collapse. The result that the ejecta 'choking' seen in the (001) case limits both the extreme KE and PE seen in the (100) shock case provides the notion that the PE increase, similar to the KE increase, is the result of the rapid recompression of expanded material. In Section 6, we will compare results for TATB to previous results for HMX[16] to assess TATB's general efficiency at localizing energy and whether the "hotspot problem" has any direct implications to TATB's insensitivity.

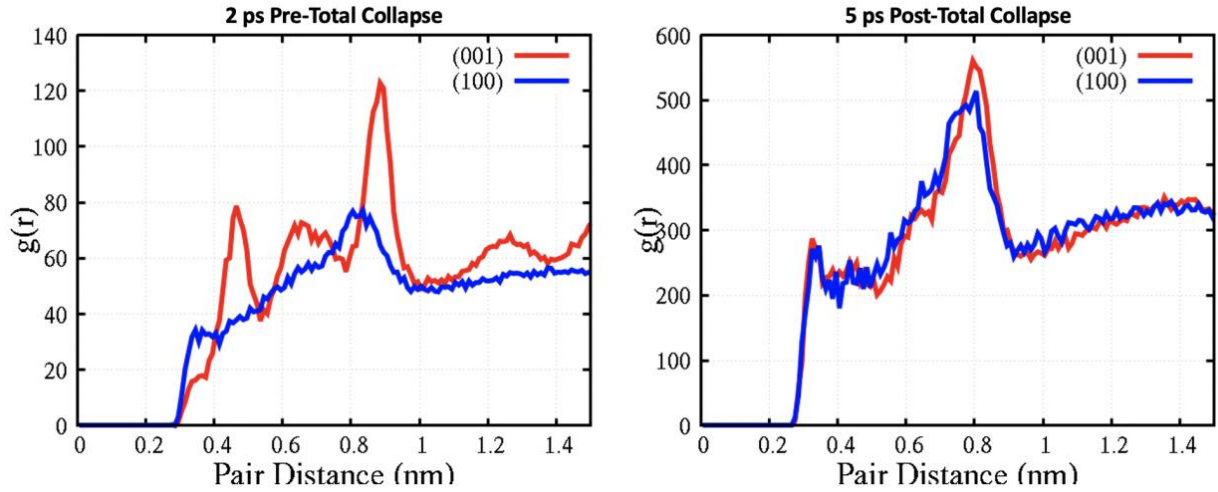

*Figure 7: Radial distribution functions for the pore collapse (mid-collapse) and the hotspot, for both orientations, cylindrical void, at 2.0 km/s.*

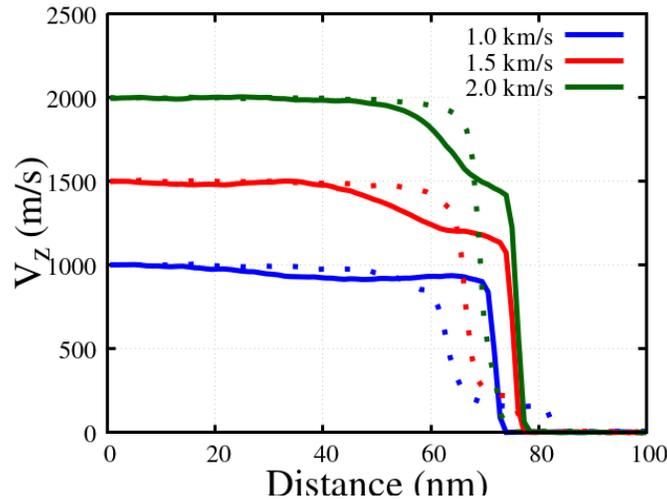

*Figure 8: Wave profiles in the bulk system for each impact velocity, dashed lines represent the (100) orientation, solid lines the (001) orientation. Both systems have a 'two-wave' feature, with the (100) leading wave causing particle velocities on the order of 200 m/s, whereas the (001) leading wave leads to significant velocity (and therefore pressure) increase.*

## 5. PE-T Distributions

Reference 45 established that hotspots are not fully described by their KE fields; the energy localized as PE cannot be inferred from the KE and the mapping is not one-to-one. The hotspots analyzed in Section 3 show a wide range of PE and temperature states for various pore shapes and shock strengths. In this section, we quantify the relationship between PE and temperature for all the cases studied to assess the role of shock strength, orientation, and defect type on this recently identified mechanism for energy localization.

Figure 9 displays scatter plots of local PE vs. temperature for the various hotspots right after total collapse of the void ($t_o + 1$ ps). These plots are broken into four subsets based on shock orientation and defect shape. The data represents a snapshot in time of the entire system, and thus show the unshocked state (low PE and temperature), the shocked states at a range of times behind the leading wave, and the hotspot. Our previous work[45] revealed that a given temperature could have a wide range of associated potential energies as different regions in the hotspots exhibit different amounts of intra-molecular strain.

As expected, the total PE and temperature are lower for weaker shocks, as is the spread of PE states for a given temperature. We find a strikingly broad distribution of potential energies in the case of circular pores while collapsed cracks show a simpler relationship between PE and T. In the case of the cylindrical pores, the spread of PE states for a given temperature is most noticeable at mid temperatures (>800 K for $U_p$ = 2.0 km/s). These results are consistent with Reference 45 that showed temperature alone cannot fully describe the thermodynamic state of the hotspot. The present results show that this observation is not specific to a single shock orientation or shock strength but a quite general feature of hotspots formed by the collapse of porosity.

Quite interestingly, (001) shocks result in two distinct PE branches while the (100) shocks exhibit a high PE hump at intermediate temperatures. The (001) results are consistent with those in Ref. 45 but the two branches were less notable in the prior work (see Fig. 4b in Ref. 45) due to coarser averaging. In order to understand the processes that result in the high-PE states, corresponding to highly deformed molecules, we map the molecules corresponding to the high-PE branch in the 2 km/s (001) show and the hump in the 2 km/s (100) into real space, see Figure 1. Our results show that, in both cases, the high-PE states do not correspond spatially the impact plane where the expanding material impacts with the downstream face of the pore. They correspond to the areas right upstream from the impact plane, with the highest plastic deformation. This observation explains why the cracks do not show high PE states for a given KE, as little plastic flow is needed to fill the void space.

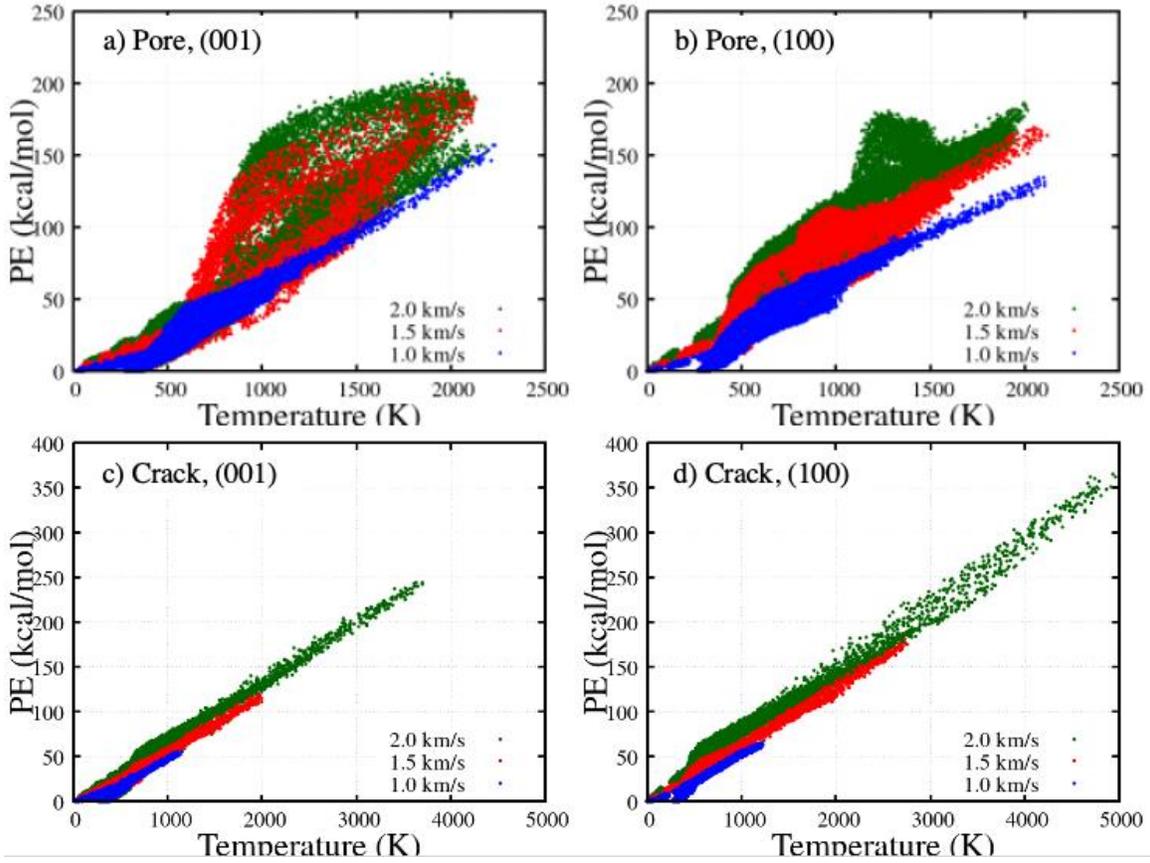

*Figure 9: PE – Temperature plots for all the hotspots for all 4 orientation/shape cases. The labels on each plot designate the defect shape, shock plane. Color designates shock strength. Distributions taken at $t_o$ + 1.0 ps.*

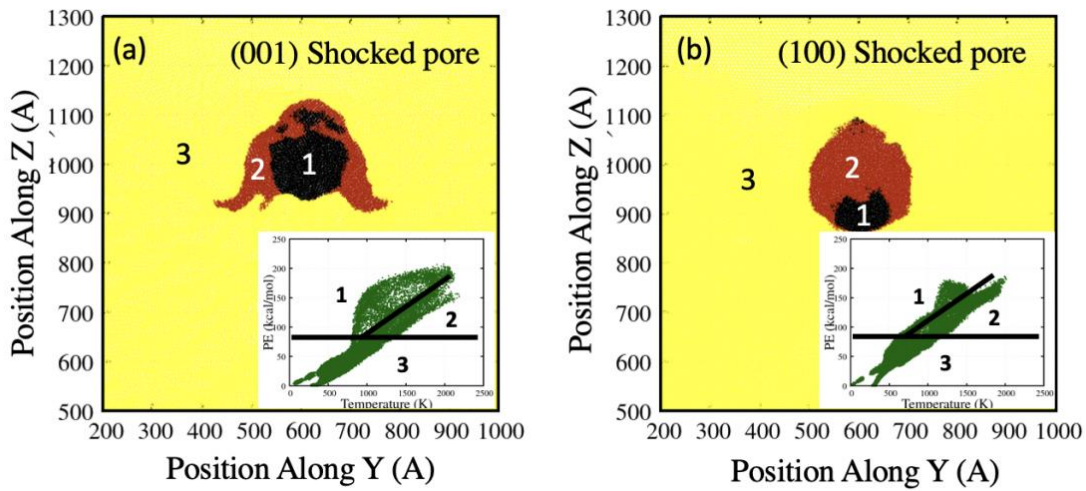

*Figure 10: Spatial location of the molecules of the 2.0 km/s pore collapses. Insets show the PE(T) distributions from Figure 9, with 3 regions depicted that are colored approapriately in the spatial plot with region 1 in the black, region 2 in red, and region 3 in yellow.*

## 6. Energy Localization Efficiency: TATB vs HMX

Figure 11 compares the temperature fields (T vs. cumulative area plots) resulting from the collapse of pores and cracks in TATB and HMX for 2.0 km/s; with HMX results for the 40nm crack and pore from Ref. 16. For all cases the area is scaled by the initial defect area for the crack and pore and we scale the temperature rise ($T_{shock} - T_{bulk}$) by the theoretical maximum hotspot temperature from Ref 15: $k_B \Delta T = \frac{m}{3} U_s U_p$ where m is the mass of the molecule. As discussed above, TATB pores and cracks have similar efficiency at localizing energy, except for a (100) shock interacting with a crack elongated along the shock direction, where we find higher temperatures. The collapse of pores in HMX result in hotspots with similar temperature distribution as in TATB. However, stronger jetting in the case of HMX cracks results in significantly higher temperatures than in TATB.

Using the cohesive energy argument from Ref 15 that was discussed in Section 4, we can compare $E_{coh}$ / MW (where MW is molecular weight) as an assessment of each material's propensity to jet. For TATB and HMX respectively, this value is 6.7 x $10^{-3}$ and 6.9 x $10^{-3}$ (eV/molecule), or more coherently, the necessary $U_p$ to jet is 1.14 and 1.15 km/s for TATB and HMX, respectively. However, the HMX crack localizes more relative energy than TATB for either orientation, supporting the notion that jetting, which may be key to high temperature hotspots seen experimentally[53–55], is the result of not just the properties of the crystal, but complex microstructural phenomena related to crystal defects and stress relaxation mechanisms. Figure 12 shows the distributions of molecular center of mass velocities for jetted molecules in both 2.0 km/s crack cases and an HMX 40nm crack from Ref. 16. In general, the molecular velocities of HMX are significantly higher than either of the TATB orientations. Despite similar $E_{coh}$ / MW values, jetted HMX molecules possess much higher KE for equivalent shock and defect conditions. These results widen a few questions regarding TATB: Do TATB crystalline defects and shear bands alleviate more energy in the bulk than HMX, causing less violent pore collapse and weaker hotspots? HMX has a higher shock speed than TATB for a given impact velocity (same input energy) and a larger slope in the $U_s$-$U_p$ relationship. Here we show a lower efficiency in generating hotspots in TATB relative to HMX, which may help to account for insensitivity to shock initiation, which is typically rationalized by molecular and chemical traits such as covalent clustering reactions[70,71] and its strong hydrogen bonding network[30,72]. The overall mechanisms behind molecular jetting and massive hotspot temperatures are still not fully understood, but obviously play a significant role in the criticality of hotspots and the overall thermo-mechanical response of a material under shock loading.

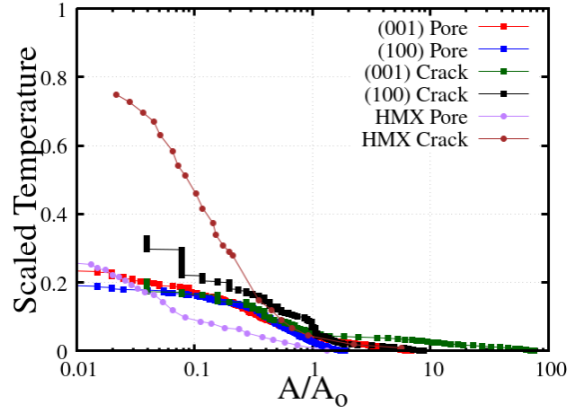

*Figure 11: Scaled Temperature-Area plots for shocks, and 2 HMX shocks from Ref 16.*

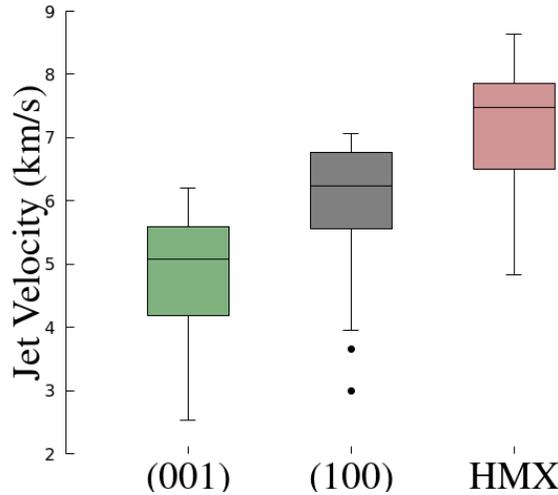

*Figure 12: Box and whisker plot of the velocities of ejected molecules for the (001) crack, (100) crack, and an HMX crack, all at a particle velocity of 2.0 km/s.*

## 7. Conclusions

In summary, we have shown that the greater localization of intra-molecular potential energy (PE) than kinetic energy (KE) occurs in TATB for a variety of impact velocities, defect shapes, and crystallographic orientations. This shows that the results of Ref. 45 are a more general finding for a variety of shock states. Within each orientation, the trends seen in PE are nearly the same as those seen in temperature. Between orientations, the difference in hotspot shape and size can be broadly explained by the molecular and crystal-level processes occurring during the collapse. For cylindrical pores, both hotspots reach similar temperatures and total areas. However, for cracks, while both orientations lead to significant molecular spall, an orientation dependence to molecular spall for crack-like defects is evident. We show this results from the shockwave progressing past the far end of the crack tip prior to spallation in some cases, which causes lateral compression of the crack that chokes off the jetted material and limits the maximum hotspot temperature. For all strong shock cases in pores ($U_p > 1.5$ km/s), there is no direct mapping between the KE and PE of the hotspot, implying that the thermodynamic state of the hotspot cannot be characterized by temperature alone. Lastly, we compare scaled hotspot temperatures to that of HMX, showing the inefficiency of TATB at creating high-temperature hotspots. This result opens new questions about the general role of crystal-level defect formation in forming hotspots, how microstructure and crystallography affect shock focusing at defects, and whether TATB's colder hotspots can be linked to its insensitivity to shock initiation.

## Acknowledgements

This work was supported by the Laboratory Directed Research and Development Program at Lawrence Livermore National Laboratory, project 18-SI-004 with Lara Leininger as P.I. Partial support was received from the US Office of Naval Research, Multidisciplinary University Research Initiatives (MURI) Program, Contract: N00014-16-1-2557. Program managers: Chad Stoltz and Kenny Lipkowitz. Simulations were made possible by computing time granted to MPK through the LLNL Computing Grand Challenge, which is gratefully acknowledged. This work was performed under the auspices of the U.S. Department of Energy by Lawrence Livermore National Laboratory under Contract DE-AC52-07NA27344. It has been approved for unlimited release under document number LLNL-JRNL-827022-DRAFT.

## References

(1) Reed, E. J.; Manaa, M. R.; Fried, L. E.; Glaesemann, K. R.; Joannopoulos, J. D. A Transient Semimetallic Layer in Detonating Nitromethane. *Nat. Phys.* **2008**, *4* (1), 72–76. https://doi.org/10.1038/nphys806.

(2) Bastea, S. Nanocarbon Condensation in Detonation. *Sci. Rep.* **2017**, *7* (November 2016), 1–6. https://doi.org/10.1038/srep42151.

(3) Reed, E. J.; Rodriguez, A. W.; Manaa, M. R.; Fried, L. E.; Tarver, C. M. Ultrafast Detonation of Hydrazoic Acid (HN 3). *Phys. Rev. Lett.* **2012**, *109* (3), 1–5. https://doi.org/10.1103/PhysRevLett.109.038301.

(4) Goldman, N.; Tamblyn, I. Prebiotic Chemistry within a Simple Impacting Icy Mixture. *J. Phys. Chem. A* **2013**, *117* (24), 5124–5131. https://doi.org/10.1021/jp402976n.

(5) Goldman, N.; Reed, E. J.; Fried, L. E.; William Kuo, I. F.; Maiti, A. Synthesis of Glycine-Containing Complexes in Impacts of Comets on Early Earth. *Nat. Chem.* **2010**, *2* (11), 949–954. https://doi.org/10.1038/nchem.827.

(6) Martins, Z.; Price, M. C.; Goldman, N.; Sephton, M. A.; Burchell, M. J. Shock Synthesis of Amino Acids from Impacting Cometary and Icy Planet Surface Analogues. *Nat. Geosci.* **2013**, *6* (12), 1045–1049. https://doi.org/10.1038/ngeo1930.

(7) Kroonblawd, M. P.; Lindsey, R. K.; Goldman, N. Synthesis of Functionalized Nitrogen-Containing Polycyclic Aromatic Hydrocarbons and Other Prebiotic Compounds in Impacting Glycine Solutions. *Chem. Sci.* **2019**, *10* (24), 6091–6098. https://doi.org/10.1039/c9sc00155g.

(8) Steele, B. A.; Goldman, N.; Kuo, I. F. W.; Kroonblawd, M. P. Mechanochemical Synthesis of Glycine Oligomers in a Virtual Rotational Diamond Anvil Cell. *Chem. Sci.* **2020**, *11* (30), 7760–7771. https://doi.org/10.1039/d0sc00755b.

(9) Shen, Y.; Reed, E. J. Quantum Nuclear Effects in Stishovite Crystallization in Shock-Compressed Fused Silica. *J. Phys. Chem. C* **2016**, *120* (31), 17759–17766. https://doi.org/10.1021/acs.jpcc.6b05083.

(10) Shen, Y.; Jester, S. B.; Qi, T.; Reed, E. J. Nanosecond Homogeneous Nucleation and Crystal Growth in Shock-Compressed SiO2. *Nat. Mater.* **2016**, *15* (1), 60–65. https://doi.org/10.1038/nmat4447.

(11) Armstrong, M. R.; Lindsey, R. K.; Goldman, N.; Nielsen, M. H.; Stavrou, E.; Fried, L. E.; Zaug, J. M.; Bastea, S. Ultrafast Shock Synthesis of Nanocarbon from a Liquid Precursor. *Nat. Commun.* **2020**, *11* (1), 1–7. https://doi.org/10.1038/s41467-019-14034-z.

(12) Hamilton, B. W.; Sakano, M. N.; Li, C.; Strachan, A. Chemistry Under Shock Conditions. *Annu. Rev. Mater. Res.* **2021**, *51* (1), 101–130. https://doi.org/10.1146/annurev-matsci-080819-120123.

(13) Campbell, A. W.; Travis, J. R. The Shock Desensitization of Pbx-9404 and Composition B-3. *Los Alamos Natl. Lab.* **1985**, No. LA-UR-85-114.

(14) Dattelbaum, D. M.; Sheffield, S. A.; Stahl, D. B.; Dattelbaum, A. M.; Trott, W.; Engelke, R. Influence of Hot Spot Features on the Initiation Characteristics of Heterogeneous Nitromethane. *Int. Detonation Symp.* **2010**, *LA-UR-10-0*.

(15) Holian, B. L.; Germann, T. C.; Maillet, J. B.; White, C. T. Atomistic Mechanism for Hot Spot Initiation. *Phys. Rev. Lett.* **2002**, *89* (28), 1–4. https://doi.org/10.1103/PhysRevLett.89.285501.

(16) Li, C.; Hamilton, B. W.; Strachan, A. Hotspot Formation Due to Shock-Induced Pore Collapse in 1,3,5,7-Tetranitro-1,3,5,7-Tetrazoctane (HMX): Role of Pore Shape and Shock Strength in Collapse Mechanism and Temperature. *J. Appl. Phys.* **2020**, *127* (17). https://doi.org/10.1063/5.0005872.

(17) Long, Y.; Chen, J. Theoretical Study of the Defect Evolution for Molecular Crystal under Shock Loading. *J. Appl. Phys.* **2019**, *125* (6). https://doi.org/10.1063/1.5067284.

(18) Springer, H. K.; Bastea, S.; Nichols, A. L.; Tarver, C. M.; Reaugh, J. E. Modeling The Effects of Shock Pressure and Pore Morphology on Hot Spot Mechanisms in HMX. *Propellants, Explos. Pyrotech.* **2018**, *43* (8), 805–817. https://doi.org/10.1002/prep.201800082.

(19) Austin, R. A.; Barton, N. R.; Reaugh, J. E.; Fried, L. E. Direct Numerical Simulation of Shear Localization and Decomposition Reactions in Shock-Loaded HMX Crystal. *J. Appl. Phys.* **2015**, *117* (18). https://doi.org/10.1063/1.4918538.

(20) Rai, N. K.; Sen, O.; Udaykumar, H. S. Macro-Scale Sensitivity through Meso-Scale Hotspot Dynamics in Porous Energetic Materials: Comparing the Shock Response of 1,3,5-Triamino-2,4,6-Trinitrobenzene (TATB) and 1,3,5,7-Tetranitro-1,3,5,7-Tetrazoctane (HMX). *J. Appl. Phys.* **2020**, *128* (8), 1–24. https://doi.org/10.1063/5.0010492.

(21) Zhao, P.; Lee, S.; Sewell, T.; Udaykumar, H. S. Tandem Molecular Dynamics and Continuum Studies of Shock-Induced Pore Collapse in TATB. *Propellants, Explos. Pyrotech.* **2020**, *45* (2), 196–222. https://doi.org/10.1002/prep.201900382.

(22) Duarte, C. A.; Li, C.; Hamilton, B. W.; Strachan, A.; Koslowski, M. Continuum and Molecular Dynamics Simulations of Pore Collapse in Shocked β-Tetramethylene Tetranitramine (β-HMX) Single Crystals. *J. Appl. Phys.* **2021**, *129* (1), 015904. https://doi.org/10.1063/5.0025050.

(23) Wood, M. A.; Kittell, D. E.; Yarrington, C. D.; Thompson, A. P. Multiscale Modeling of Shock Wave Localization in Porous Energetic Material. *Phys. Rev. B* **2018**, *97* (1), 1–9. https://doi.org/10.1103/PhysRevB.97.014109.

(24) Strachan, A.; van Duin, A. C. T.; Chakraborty, D.; Dasgupta, S.; Goddard, W. A. Shock Waves in High-Energy Materials: The Initial Chemical Events in Nitramine RDX. *Phys. Rev. Lett.* **2003**, *91* (9), 7–10. https://doi.org/10.1103/PhysRevLett.91.098301.

(25) Strachan, A.; Kober, E. M.; Van Duin, A. C. T.; Oxgaard, J.; Goddard, W. A. Thermal Decomposition of RDX from Reactive Molecular Dynamics. *J. Chem. Phys.* **2005**, *122* (5). https://doi.org/10.1063/1.1831277.

(26) Zybin, S. V.; Goddard, W. A.; Xu, P.; Van Duin, A. C. T.; Thompson, A. P. Physical Mechanism of Anisotropic Sensitivity in Pentaerythritol Tetranitrate from Compressive-Shear Reaction Dynamics Simulations. *Appl. Phys. Lett.* **2010**, *96* (8), 1–4. https://doi.org/10.1063/1.3323103.

(27) Guo, D.; An, Q.; Goddard, W. A.; Zybin, S. V.; Huang, F. Compressive Shear Reactive Molecular Dynamics Studies Indicating That Cocrystals of TNT/CL-20 Decrease Sensitivity. *J. Phys. Chem. C* **2014**, *118* (51), 30202–30208. https://doi.org/10.1021/jp5093527.

(28) Riad Manaa, M.; Reed, E. J.; Fried, L. E. Atomistic Simulations of Chemical Reactivity of TATB under Thermal and Shock Conditions. *Proc. - 14th Int. Detonation Symp. IDS 2010* **2010**, 837–843.

(29) Manaa, M. R.; Goldman, N.; Fried, L. E. Simulation of Cluster Formation in Overdriven and Expanding TATB by Molecular Dynamics. *Proc. 15th Int. Detonation Symp.* **2015**, 53–59.

(30) Manaa, M. R. Shear-Induced Metallization of Triamino-Trinitrobenzene Crystals. *Appl. Phys. Lett.* **2003**, *83* (7), 1352–1354. https://doi.org/10.1063/1.1603351.

(31) Hamilton, B. W.; Kroonblawd, M. P.; Islam, M. M.; Strachan, A. Sensitivity of the Shock Initiation Threshold of 1,3,5-Triamino-2,4,6-Trinitrobenzene (TATB) to Nuclear Quantum Effects. *J. Phys. Chem. C* **2019**, *123* (36), 21969–21981. https://doi.org/10.1021/acs.jpcc.9b05409.

(32) Guo, D.; Zybin, S. V.; An, Q.; Goddard, W. A.; Huang, F. Prediction of the Chapman-Jouguet Chemical Equilibrium State in a Detonation Wave from First Principles Based Reactive Molecular Dynamics. *Phys. Chem. Chem. Phys.* **2016**, *18* (3), 2015–2022. https://doi.org/10.1039/c5cp04516a.

(33) Islam, M. M.; Strachan, A. Decomposition and Reaction of Polyvinyl Nitrate under Shock

and Thermal Loading: A ReaxFF Reactive Molecular Dynamics Study. *J. Phys. Chem. C* **2017**, *121* (40), 22452–22464. https://doi.org/10.1021/acs.jpcc.7b06154.

(34) Islam, M. M.; Strachan, A. Reactive Molecular Dynamics Simulations to Investigate the Shock Response of Liquid Nitromethane. *J. Phys. Chem. C* **2019**, *123* (4), 2613–2626. https://doi.org/10.1021/acs.jpcc.8b11324.

(35) Powell, M. S.; Sakano, M. N.; Cawkwell, M. J.; Bowlan, P. R.; Brown, K. E.; Bolme, C. A.; Moore, D. S.; Son, S. F.; Strachan, A.; McGrane, S. D. Insight into the Chemistry of PETN under Shock Compression through Ultrafast Broadband Mid-Infrared Absorption Spectroscopy. *J. Phys. Chem. A* **2020**, *124* (35), 7031–7046. https://doi.org/10.1021/acs.jpca.0c03917.

(36) Zhang, L.; Zybin, S. V; Duin, A. C. T. Van; Dasgupta, S.; Iii, W. A. G.; Kober, E. M. High Explosives from ReaxFF Reactive Molecular Dynamics Simulations. *Society* **2009**, 10619–10640.

(37) Hamilton, B. W.; Steele, B. A.; Sakano, M. N.; Kroonblawd, M. P.; Kuo, I. F. W.; Strachan, A. Predicted Reaction Mechanisms, Product Speciation, Kinetics, and Detonation Properties of the Insensitive Explosive 2,6-Diamino-3,5-Dinitropyrazine-1-Oxide (LLM-105). *J. Phys. Chem. A* **2021**, *125* (8), 1766–1777. https://doi.org/10.1021/acs.jpca.0c10946.

(38) Wu, C. J.; Fried, L. E. Ab Initio Study of RDX Decomposition Mechanisms. *J. Phys. Chem. A* **1997**, *101* (46), 8675–8679. https://doi.org/10.1021/jp970678+.

(39) Wu, C. J.; Fried, L. E. Ring Closure Mediated by Intramolecular Hydrogen Transfer in the Decomposition of a Push-Pull Nitroaromatic: TATB. *J. Phys. Chem. A* **2000**, *104* (27), 6447–6452. https://doi.org/10.1021/jp001019r.

(40) Sakano, M.; Hamilton, B. W.; Islam, M. M.; Strachan, A. Role of Molecular Disorder on the Reactivity of RDX. *J. Phys. Chem. C* **2018**, *122* (47), 27032–27043. https://doi.org/10.1021/acs.jpcc.8b06509.

(41) Sakano, M. N.; Hamed, A.; Kober, E. M.; Grilli, N.; Hamilton, B. W.; Islam, M. M.; Koslowski, M.; Strachan, A. Unsupervised Learning-Based Multiscale Model of Thermochemistry in 1,3,5-Trinitro-1,3,5-Triazinane (RDX). *J. Phys. Chem. A* **2020**, *124* (44), 9141–9155. https://doi.org/10.1021/acs.jpca.0c07320.

(42) Wood, M. A.; Cherukara, M. J.; Kober, E. M.; Strachan, A. Ultrafast Chemistry under Nonequilibrium Conditions and the Shock to Deflagration Transition at the Nanoscale. *J. Phys. Chem. C* **2015**, *119* (38), 22008–22015. https://doi.org/10.1021/acs.jpcc.5b05362.

(43) Shan, T. R.; Wixom, R. R.; Thompson, A. P. Extended Asymmetric Hot Region Formation Due to Shockwave Interactions Following Void Collapse in Shocked High Explosive. *Phys. Rev. B* **2016**, *94* (5). https://doi.org/10.1103/PhysRevB.94.054308.

(44) Kroonblawd, M. P.; Fried, L. E. High Explosive Ignition through Chemically Activated Nanoscale Shear Bands. *Phys. Rev. Lett.* **2020**, *124* (20), 206002. https://doi.org/10.1103/PhysRevLett.124.206002.

(45) Hamilton, B. W.; Kroonblawd, M. P.; Li, C.; Strachan, A. A Hotspot's Better Half: Non-

Equilibrium Intra-Molecular Strain in Shock Physics. *J. Phys. Chem. Lett.* **2021**, *12* (11), 2756–2762. https://doi.org/10.1021/acs.jpclett.1c00233.

(46) Ribas-Arino, J.; Marx, D. Covalent Mechanochemistry: Theoretical Concepts and Computational Tools with Applications to Molecular Nanomechanics. *Chem. Rev.* **2012**, *112* (10), 5412–5487. https://doi.org/10.1021/cr200399q.

(47) Islam, M. M.; Strachan, A. Role of Dynamical Compressive and Shear Loading on Hotspot Criticality in RDX via Reactive Molecular Dynamics. *J. Appl. Phys.* **2020**, *128* (6). https://doi.org/10.1063/5.0014461.

(48) Cady, H. H.; Larson, A. C. The Crystal Structure of 1,3,5-Triamino-2,4,6-Trinitrobenzene. *Acta Crystallogr.* **1965**, *18* (3), 485–496. https://doi.org/10.1107/s0365110x6500107x.

(49) Lafourcade, P.; Denoual, C.; Maillet, J. B. Irreversible Deformation Mechanisms for 1,3,5-Triamino-2,4,6-Trinitrobenzene Single Crystal through Molecular Dynamics Simulations. *J. Phys. Chem. C* **2018**, *122* (26), 14954–14964. https://doi.org/10.1021/acs.jpcc.8b02983.

(50) Zhao, P.; Kroonblawd, M. P.; Mathew, N.; Sewell, T. Strongly Anisotropic Thermomechanical Response to Shock Wave Loading in Oriented Samples of the Triclinic Molecular Crystal 1,3,5-Triamino-2,4,6-Trinitrobenzenen (TATB). *chemrxiv* **2021**.

(51) Kroonblawd, M. P.; Sewell, T. D. Theoretical Determination of Anisotropic Thermal Conductivity for Initially Defect-Free and Defective TATB Single Crystals. *J. Chem. Phys.* **2014**, *141* (18). https://doi.org/10.1063/1.4901206.

(52) Steele, B. A.; Clarke, S. M.; Kroonblawd, M. P.; Kuo, I. F. W.; Pagoria, P. F.; Tkachev, S. N.; Smith, J. S.; Bastea, S.; Fried, L. E.; Zaug, J. M.; Stavrou, E.; Tschauner, O. Pressure-Induced Phase Transition in 1,3,5-Triamino-2,4,6-Trinitrobenzene (TATB). *Appl. Phys. Lett.* **2019**, *114* (19). https://doi.org/10.1063/1.5091947.

(53) Bassett, W. P.; Johnson, B. P.; Neelakantan, N. K.; Suslick, K. S.; Dlott, D. D. Shock Initiation of Explosives: High Temperature Hot Spots Explained. *Appl. Phys. Lett.* **2017**, *111* (6). https://doi.org/10.1063/1.4985593.

(54) Bassett, W. P.; Johnson, B. P.; Dlott, D. D. Dynamic Absorption in Optical Pyrometry of Hot Spots in Plastic-Bonded Triaminotrinitrobenzene. *Appl. Phys. Lett.* **2019**, *114* (19). https://doi.org/10.1063/1.5092984.

(55) Bassett, W. P.; Johnson, B. P.; Salvati, L.; Nissen, E. J.; Bhowmick, M.; Dlott, D. D. Shock Initiation Microscopy with High Time and Space Resolution. *Propellants, Explos. Pyrotech.* **2020**, *45* (2), 223–235. https://doi.org/10.1002/prep.201900222.

(56) Plimpton, S. Fast Parallel Algorithms for Short-Range Molecular Dynamics. *Journal of Computational Physics*. 1995, pp 1–19. https://doi.org/10.1006/jcph.1995.1039.

(57) Bedrov, D.; Borodin, O.; Smith, G. D.; Sewell, T. D.; Dattelbaum, D. M.; Stevens, L. L. A Molecular Dynamics Simulation Study of Crystalline 1,3,5-Triamino-2,4,6- Trinitrobenzene as a Function of Pressure and Temperature. *J. Chem. Phys.* **2009**, *131* (22). https://doi.org/10.1063/1.3264972.

(58) Kroonblawd, M. P.; Sewell, T. D. Theoretical Determination of Anisotropic Thermal Conductivity for Crystalline 1,3,5-Triamino-2,4,6-Trinitrobenzene (TATB). *J. Chem. Phys.* **2013**, *139* (7). https://doi.org/10.1063/1.4816667.

(59) Andersen, H. C. Rattle: A "Velocity" Version of the Shake Algorithm for Molecular Dynamics Calculations. *J. Comput. Phys.* **1983**, *52* (1), 24–34. https://doi.org/10.1016/0021-9991(83)90014-1.

(60) Mathew, N.; Sewell, T. D.; Thompson, D. L. Anisotropy in Surface-Initiated Melting of the Triclinic Molecular Crystal 1,3,5-Triamino-2,4,6-Trinitrobenzene: A Molecular Dynamics Study. *J. Chem. Phys.* **2015**, *143* (9). https://doi.org/10.1063/1.4929806.

(61) Wolf, D.; Keblinski, P.; Phillpot, S. R.; Eggebrecht, J. Exact Method for the Simulation of Coulombic Systems by Spherically Truncated, Pairwise r-1 Summation. *J. Chem. Phys.* **1999**, *110* (17), 8254–8282. https://doi.org/10.1063/1.478738.

(62) Kroonblawd, M. P.; Mathew, N.; Jiang, S.; Sewell, T. D. A Generalized Crystal-Cutting Method for Modeling Arbitrarily Oriented Crystals in 3D Periodic Simulation Cells with Applications to Crystal–Crystal Interfaces. *Comput. Phys. Commun.* **2016**, *207*, 232–242. https://doi.org/10.1016/j.cpc.2016.07.007.

(63) Nosé, S. A Unified Formulation of the Constant Temperature Molecular Dynamics Methods. *J. Chem. Phys.* **1984**, *81* (1), 511–519. https://doi.org/10.1063/1.447334.

(64) Holian, B. L.; Lomdahl, P. S. Plasticity Induced by Shock Waves in Nonequilibrium Molecular-Dynamics Simulations. *Science (80-. ).* **1998**, *280* (5372), 2085–2088. https://doi.org/10.1126/science.280.5372.2085.

(65) Tarver, C. M.; Chidester, S. K.; Nichols, A. L. Critical Conditions for Impact- and Shock-Induced Hot Spots in Solid Explosives. *J. Phys. Chem.* **1996**, *100* (14), 5794–5799. https://doi.org/10.1021/jp953123s.

(66) Kroonblawd, M. P.; Hamilton, B. W.; Strachan, A. Fourier-like Thermal Relaxation of Nanoscale Explosive Hot Spots. *J. Phys. Chem. C* **2021**, *125*, 20570–20582. https://doi.org/10.1021/acs.jpcc.1c05599.

(67) Taw, M. R.; Yeager, J. D.; Hooks, D. E.; Carvajal, T. M.; Bahr, D. F. The Mechanical Properties of As-Grown Noncubic Organic Molecular Crystals Assessed by Nanoindentation. *J. Mater. Res.* **2017**, *32* (14), 2728–2737. https://doi.org/10.1557/jmr.2017.219.

(68) Pal, A.; Picu, C. R. Non-Schmid Effect of Pressure on Plastic Deformation in Molecular Crystal HMX. *J. Appl. Phys.* **2019**, *125* (21), 1–9. https://doi.org/10.1063/1.5092285.

(69) Forbes, J. W. *Shock Wave Compression of Condensed Matter*; Springer Science & Business Media, 2013.

(70) Manaa, M. R.; Reed, E. J.; Fried, L. E.; Goldman, N. Nitrogen-Rich Heterocycles as Reactivity Retardants in Shocked Insensitive Explosives. *J. Am. Chem. Soc.* **2009**, *131* (15), 5483–5487. https://doi.org/10.1021/ja808196e.

(71) Tiwari, S. C.; Nomura, K. I.; Kalia, R. K.; Nakano, A.; Vashishta, P. Multiple Reaction

Pathways in Shocked 2,4,6-Triamino-1,3,5-Trinitrobenzene Crystal. *J. Phys. Chem. C* **2017**, *121* (29), 16029–16034. https://doi.org/10.1021/acs.jpcc.7b05253.

(72) Manaa, M. R.; Gee, R. H.; Fried, L. E. Internal Rotation of Amino and Nitro Groups in TATB: MP2 versus DFT (B3LYP). *J. Phys. Chem. A* **2002**, *106* (37), 8806–8810. https://doi.org/10.1021/jp0259972.